\documentclass{epl}
\input{psfig.sty}

\title{Maximum matching on random graphs} 

\author{Haijun Zhou\inst{1,2,3} \and  Zhong-can Ou-Yang\inst{1}}
\institute{
  \inst{1} Institute of Theoretical Physics, the Chinese 
  Academy of Sciences,
  P.O. Box 2735, Beijing 100080, China \\
  \inst{2} Interdisciplinary Center of Theoretical Studies, the Chinese
  Academy of Sciences, P.O. Box 2735, Beijing 100080, China \\
  \inst{3} Max-Planck-Institute of Colloids and Interfaces, 
  14424 Potsdam, Germany 
} 

\begin{document}

\maketitle

\begin{abstract}
  The maximum matching problem on random graphs is studied 
  analytically by  the cavity method of 
  statistical physics. When the average vertex degree \mth{c} is 
  larger than \mth{2.7183}, groups of max-matching patterns  which 
  differ greatly from each other {\em gradually} emerge.  
  An analytical expression for the max-matching size is also 
  obtained, which agrees well with computer simulations. 
  Discussion is made on this {\em continuous} glassy phase transition 
  and the absence of such a glassy phase in the related
  minimum vertex covering problem.
\end{abstract}

\section{Introduction}

Studies on spin glasses focus on systems with random frustrations 
\cite{Edwards1975,Binder1986}. The energy landscape of a spin glass 
is very rough. When environmental temperature is lower than certain
critical value, the system gets trapped in one of many local regions of the
whole configurational space (ergodicity breaking). 
The deep connection
between frustrations in spin glasses and constraints in combinatorial
optimization problems was noticed by many authors,
and the replica method  developed during
the study of spin glass physics \cite{Binder1986,mezard1987} has 
been applied to hard combinatorial optimization problems including 
the k-satisfiability \cite{monasson1997},
the number partitioning \cite{mertens1998}, the Euclidean matching
\cite{houdayer1998}, the vertex covering \cite{weigt2001},
and many others. However, sophistication of replica
method renders analytical discussion to be limited usually to the 
replica-symmetric (RS) level.

Recently, considerable success has been attained in applying the 
cavity method \cite{mezard2001,mezard2003} to combinatorial
optimization problems \cite{mezard2002,mulet2002,mezard2003b,zhou2003b}.
The cavity formalism enables analytical calculations to be 
carried out to first-step replica-symmetry-breaking (RSB). 
For random 3-satisfiability (3-Sat) problems it was discovered
\cite{mezard2002} that, between the Easy-SAT and UNSAT phase there
is a Hard-SAT phase. The Hard-SAT phase is a glassy phase, with great 
many states of the same ground-state energy density which 
are separated by very high energy barriers. The Easy-SAT to Hard-SAT
phase-transition is abrupt. Similar behavior was observed in
random 3-XOR-Sat \cite{mezard2003b}. On the other hand, work performed
on minimum vertex covering (min-covering) of random 
graphs \cite{zhou2003b} suggested
there is no proliferation of ground-states in this model system
even when replica symmetry is broken. 

The following questions arise:  (1) Why in the RSB domain, 
the structural entropy density (complexity) of min-covering is zero?
(2) If there is a glassy phase-transition in a combinatorial optimization
problem, is it always discontinuous? In this article, we
looked into these questions by studying the maximum matching (max-matching)
problem of random graphs. We chose to work with max-matching because
(a) as will be shown later, max-matching is equivalent to
min-covering in the RS parameter range and,
(b) there exist polynomial algorithms to
find a max-matching, so that theory can be checked by
experiment.

We found that, when  the average number of 
nearest-neighboring vertices of each vertex in  a random graph, 
called the average vertex degree $c$,
is larger than $c_{\rm cr}=e$ the system is in a
glassy phase with many max-matchings that have very large
Hamming distance between each other. Contrast to the situations
in random 3-Sat and 3-XOR-Sat, these  patterns appear gradually;
their number increases exponentially with $c$. 
When $c<e$, max-matching and min-covering are equivalent 
(there is no edge-redundancy). 
When $c\geq e$, the appearance of redundant edges, while adding 
great freedom to max-matching, cause severe constraints
to min-covering patterns, since each edge needs to be covered in 
the later case.

We also obtained an analytical expression for the average max-match
size, which is in agreement with computer simulations in the whole
range of parameter $c$.

\section{Model and cavity calculations}
Consider a random graph $G(n,c/(n-1))$ \cite{BollobasB1985}. 
There are $n$ vertices in the
vertex set $V=\{v_1,v_2,\ldots,v_n\}$; between any pair of 
vertices $v_i$ and $v_j$,
an edge is present in the edge set $E$ with probability
$c/(n-1)$ and absent with probability $1-c/(n-1)$. The average
number of edges incident to a randomly chosen vertex is $c$; 
and for large graph size $n$, the number $k$ of edges associated
with a given vertex obeys the Poisson distribution 
$P_{P}(c,k)=e^{-c} c^k /k!$ \cite{BollobasB1985}. 
A  matching $M$ is a subset of the edge set $E$ such
that no two edges in $M$ share a common vertex. We are interested
in the {\em max-matching} $M^*$, a matching with the largest number 
of edges. What is the size $|M^*|$ of a max-matching for
a random graph $G(n,c/(n-1))$ constructed by the above mentioned procedure?
This question could be answered analytically and algorithmically.
First, we investigate random graph max-matching by zero-temperature
statistical physics, and  an analytical formula will be given. 
The max-matching patterns could be classified into different groups 
based on their mutual similarity. It turns out that
the total number of such groups grow exponentially with graph size
$n$ when the average vertex degree $c>e$.

We associate with each {\em edge}\footnote{Notice that, different from
  ref.~\cite{zhou2003b}, spins are associated 
  with edges of the graph and  cavity fields (see below)
  are associated with vertices of the graph. }
$e_i$ of graph $G(n,c/(n-1))$ a 
spin variable $S=\{0,1\}$. There are altogether $2^{|E|}$ possible
microscopic spin configurations for the graph. We introduce the
following energy functional for each  microscopic spin configuration:
\begin{equation}
  \label{eq:energy_functional}
  {\cal{H}}[\{S\}]=-\sum\limits_{e_{i}\in E} S_{e_i}
  +\lambda \sum{^\prime} S_{e_i} S_{e_j},
\end{equation}
where $\lambda$ ($>1$) is a constant and the prime 
indicates the second summation is restricted to edges
$e_i$ and $e_j$ that share a common vertex. For a max-matching
$M^*$, we assign $S_{e_i}=1$ to all the edges 
$e_i\in M^*$ and $S_{e_j}=0$ to all the other edges $e_j$. 
The configurational energy is $-|M^*|$. Because $\lambda>1$,
no microscopic spin
configurations could attain energy lower than $-|M^*|$;
therefore, finding the max-matching size is converted to
finding the ground energy states of eq.~(\ref{eq:energy_functional}). 
Furthermore, each energy (local) minimum configuration
corresponds to a matching of the graph \cite{zhou2003b}.
Hereafter we consider only these energy minimum configurations.
For very large systems ($n\gg 1$) we  group them into different 
(macroscopic) states. A state of the system contains a set of 
microscopic spin configurations, all of which
have the same (minimum) energy and only finite 
number (or number at most proportional to $n^\theta$ with $\theta<1$)
of spin flips is needed to transit from one to another of 
these configurations \cite{mezard2001}.

For each state, say $\alpha$, we define a quantity $h_{\alpha}(v_i)$
(called the cavity field) for each {\em vertex} $v_i$ 
according to  the following rule:
$h_{\alpha}(v_i)=-1$, if in each microscopic configuration
of state $\alpha$ one of the edges that meets $v_i$ has spin value $S=1$;
and $h_{\alpha}(v_i)=0$, if in one or more microscopic 
configurations of state $\alpha$ all the edges that meet $v_i$ have spin 
value $S=0$. 

A random graph $G(n,c/(n-1))$ can be obtained by first generating  a random
graph $G(n-1,c/(n-1))$ and then adding a new vertex (say $v_0$)
and setting up $k$ edges
$e_1, e_2,\ldots,e_k$ to $k$ randomly chosen vertices 
(say $v_1$, $v_2$, \ldots,
$v_k$) of the old graph.
The value of $k$ is governed by $P_P(c,k)$ when $n$ is considerably 
large.
If the original system is in state $\alpha$,
the energy difference between the enlarged and the original system is
\begin{equation}
  \label{eq:energy_difference}
  \triangle E_{\alpha}=\min\limits_{S_{e_1},\ldots,S_{e_k}}
  \left[-\sum\limits_{i=1}^{k} [h_\alpha(v_i)+1] S_{e_i}
    +\lambda \sum\limits_{i=1}^{k-1} \sum\limits_{j=i+1}^{k}
    S_{e_i} S_{e_j}\right].
\end{equation}
For the purpose of favoring  microscopic spin configurations that
lead to an energy decrease, we introduce an $``$effective 
inverse temperature''
parameter $y$ \cite{mezard2002,mezard2003}. Denote $P_{v_i}^{(y)}(h)$
as the distribution of the cavity field $h$ of vertex $v_i$ over
different states $\alpha$ of the graph at given effective inverse temperature
$y$. The following recursive
equation could be written down for this probability distribution:
\begin{equation}
  P_{v_0}^{(y)}(h_0)=\delta(k) \delta(h_0)
  +[1-\delta(k)] C  \prod\limits_{i=1}^k [ \int \drm h_i P_{v_i}^{(y)}
  (h_i) ] e^{-y \triangle E} \delta[h_0+\theta(\sum\limits_{i=1}^k h_i +k)],
  \label{eq:h_distribution}
\end{equation}
where $\delta(x)$ is the Kronecker symbol;
$C$ is a normalization constant; and
$\theta(x)=1$ if $x>0$ and $=0$ otherwise. 
Equation (\ref{eq:h_distribution}) is understood as follows: 
(1) if vertex $v_0$ is  isolated,  it feels no field (the first term);
(2) if $k\geq 1$ but in state $\alpha$
each of $v_0$'s  neighbors has already associated with an edge of $S=1$,
the edges of $v_0$ should all assign $S=0$ to decrease
energy ($h_0=0$); (3) otherwise, it is energetically favorable to 
assign $S=1$ to  one of the edges of vertex $v_0$ ($h_0=1$). 
Situation (2) and (3) correspond to the second term of 
eq.~(\ref{eq:h_distribution}).

In deriving eq.~(\ref{eq:h_distribution}) it was assumed that, 
before the addition of vertex $v_0$, the cavity field distributions 
for the vertices $v_1$,$v_2$,\ldots,$v_k$ are mutually 
independent \cite{mezard2001,mezard2003}. 
This assumption is based on
the argument that, before $v_0$ is added these vertices  are usually
far apart. It certainly does not hold for regular networks.

A careful inspection of eqs.~(\ref{eq:energy_difference}) and
(\ref{eq:h_distribution}) leads to the following form for the cavity
field distribution:
\begin{equation}
  \label{eq:h_form}
  P^{(y)}(h)=
  \left\{
    \begin{array}{cl}
      \delta(h+1), & {\rm probability}\;\;p_1 \\
      \delta(h),   & {\rm probability}\;\;p_2\\
      \alpha \delta(h+1)+(1-\alpha) \delta(h),
      \;\; & {\rm probability}\;\;p_3
    \end{array}
  \right.
\end{equation}
where $0<\alpha<1$ is a random number obeying certain distribution;
and 
\begin{equation}
  \label{eq:p1p2}
  p_1=1-e^{-c p_2}, \;\;\;\; p_2=e^{-c (1-p_1)},
  \;\;p_3=1-p_1-p_2.
\end{equation}
We introduce the following transformation for
variable $\alpha$ in eq.~(\ref{eq:h_form}):
$\alpha=1/[1+\tau e^{-y/2}]$. It could be shown that
$\tau >0$ is governed by the population dynamics equation
\begin{eqnarray}
  \rho^{(y)}(\tau)&=&\sum\limits_{m=1}^\infty {
    P_P(c p_3, m) \over 1-e^{-c p_3}}
  \prod\limits_{i=1}^m [\int \drm \tau_i \rho^{(y)}(\tau_i) ]
  \delta\left[
    \tau-{e^{-y/2} \over \prod\limits_{j=1}^m [1+\tau_j e^{-y/2}]-1}\right]
  \label{eq:tau} \\
  &=&\sum\limits_{m=1}^\infty{
    P_P(m,c p_3) \over 1-e^{-c p_3}}
  \prod\limits_{i=1}^m [\int \drm \tau_i \rho^{(y)}(\tau_i)]
  \delta(\tau-{1\over \tau_1+\ldots+\tau_m})
  \;\;\;\;(y\rightarrow +\infty).
  \label{eq:tau_infty}
\end{eqnarray}

The free energy density at given value of the effective inverse 
temperature $y$ is obtained following the procedure given 
in ref.~\cite{mezard2002b}
(see also ref.~\cite{zhou2003b}). The expression reads
\begin{eqnarray}
  \Phi(y) &=&{1\over 2}(-1-p_1+p_2+c p_2-c p_1 p_2)
  +(p_3/y)\left[(1+c p_2) \overline{\ln(\tau+ e^{-y/2})}\right. \nonumber \\
  & &\left.+{c\over 2} p_3^2 {\overline{\ln(1+\tau_1 \tau_2 +\tau_1 e^{-y/2}
        +\tau_2 e^{-y/2})}}
    -(1+c-c p_1) {\overline{\ln(1+\tau e^{-y/2})}}\right]
  \label{eq:free_energy} \\
  &=&{1\over 2}(-1-p_1+p_2+c p_2-c p_1 p_2) \nonumber \\
  & & 
  +(p_3/y)[(1+c p_2) {\overline{\ln \tau}}
  +{c\over 2} p_3^2 {\overline{\ln(1+\tau_1 \tau_2)}}]
  \;\;\;\;(y\gg 1).
  \label{eq:free_limit}
\end{eqnarray}
In the above equation and hereafter, an overline means performing
an averaging.

We are interested in the average size of a max-matching for a
random graph $G(n,c/(n-1))$. This corresponds to the 
free energy of the system at $y\rightarrow \infty$, provided
that the complexity or structural entropy density, $\Sigma(y)=d\Phi(y)/d(1/y)$,
is non-negative (that is, if there exist states at $y\rightarrow
\infty$) \cite{mezard2002}. When the average vertex degree
$c\leq e$, the only solution
of eq.~(\ref{eq:p1p2}) is $p_1=1-W(c)/c$ and $p_2=W(c)/c$, where
$W(c)$ is the root of $We^{W}=c$. In this case 
the average size of the max-matching (in units of the graph size $n$) is
\begin{equation}
  \label{eq:rs}
  x(c)=\lim\limits_{n\rightarrow\infty} {{\overline{|M^*|}} \over n}
  =1-W(c)/c-W^2(c)/2 c.
\end{equation}
We see the average max-matching size is identical to the 
average minimum vertex
cover size obtained in refs.~\cite{weigt2001,weigt2000}. This should be
the case.
Actually, for $c\leq e$ the leaf-removal algorithm of ref.~\cite{bauer2001}
will at the same time report a min-covering and a max-matching. 
Expression (\ref{eq:rs}) is exact for $c\leq e$; while for 
$c>e$ it is just an upper-bound (for $c\geq 4$ it even exceeds $1/2$).

When the average degree $c>e$, eq.~(\ref{eq:p1p2}) has another solution
with $p_1+p_2<1$,
and eq.~(\ref{eq:free_limit}) leads to the following max-matching
size expression
\begin{equation}
  \label{eq:rsb}
  x(c)=(1+p_1-p_2-c p_2+c p_1 p_2)/2.
\end{equation}
The  structural entropy density at effective inverse temperature $y=\infty$ is
\begin{equation}
  \label{eq:entropy}
  \Sigma=p_3 (1+c p_2) {\overline{\ln \tau}}+
  (c/2) p_3^2 {\overline{\ln(1+\tau_1 \tau_2)}}.
\end{equation}

\begin{figure}[tb]
  \centerline{\psfig{file=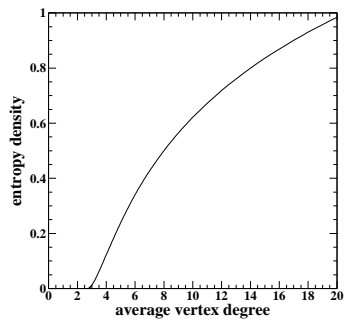,width=5.0cm}}
  \caption{
    \label{fig:fig01}
    The structural entropy density at $y=\infty$ [eq.~(\ref{eq:entropy})] as
    a function of average vertex degree $c$. The
    structural entropy density could exceed $\ln 2$, since
    the total number of configurations is $2^{c n/2}$ rather than $2^n$.
  }
\end{figure}

Based on eqs.~(\ref{eq:tau_infty}) and (\ref{eq:entropy}), the 
ground-state structural entropy density is calculated numerically for various
values of $c$ and is shown in fig.~\ref{fig:fig01}. The structural
entropy density is zero for $c\leq e$ and it gradually increases
from zero as $c>e$. Therefore, for $c>e$  there
exist many degenerate ground states and the system is in a
zero-temperature glassy phase. Figure~\ref{fig:fig01} also 
reveals that, the larger the value of $c$, 
the larger the value of the structural entropy density.
This observation is
quite different from the case of random 3-Sat  and 3-XOR-Sat
problem \cite{mezard2002,mezard2003}.
In random 3-Sat and 3-XOR-Sat problems, at the phase-transition
point, the complexity jumps from zero to a finite value
and then gradually decreases to zero as the average vertex degree $c$
increases.

The ``phase-diagram'' fig.~\ref{fig:fig01} is also quite different
from that of the minimum vertex covering problem \cite{zhou2003b}.
It is noticeable that, while the minimum vertex covering problem 
could be mapped to an energy functional very similar to 
eq.~(\ref{eq:energy_functional}) in form, in that system 
there is no zero-temperature  glassy phase \cite{zhou2003b}. We
will discuss this in the last section.

\begin{figure}[tb]
  \centerline{\psfig{file=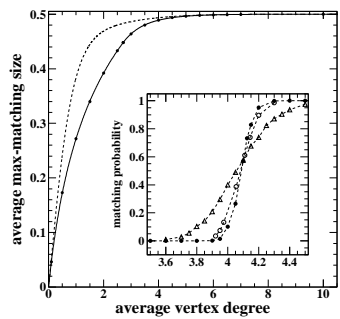,width=5.0cm}}
  \caption{
    \label{fig:fig02}
    The average size of the max-matchings for a random graph of
    fixed average vertex degree (solid line) and its asymptotic
    form (dashed line, see eq.~(\ref{eq:scaling})).
    Filled circles are average max-matching sizes obtained by a 
    matching algorithm (see appendix).
    At each given average vertex degree $c$, in
    the numerical simulation,  $1000$ samples of a random graph of 
    $10000$ vertices were generated and the max-matching size for
    each of them was obtained and then averaged. 
    The errors in the experimental
    data are all smaller than the radius of the circles.
    Inset is the probability of a random graph of  $n$ vertices
    to have a matching containing at least $0.49 n $ edges:
    $n=10000$, filled circles ($2000$ samples per point);
    $n=5000$, empty circles ($2000$ samples per point); 
    $n=1000$, empty triangles ($3000$ samples per point).
  }
\end{figure}

In fig.~\ref{fig:fig02} the average size of max-matching 
eq.~(\ref{eq:rsb}) is shown for
various values of $c$. Also shown are the results obtained by an 
exact algorithm mentioned in the appendix. 
The analytical results are in complete
agreement with the experimental result in the whole range of $c$. 
We suggest that the analytical expressions eq.~(\ref{eq:rsb}) and
(\ref{eq:entropy}) obtained by statistical mechanics method  are exact
in the limit $n\gg 1$. The scaling of max-matching size for $c\gg 1$
is
\begin{equation}
  \label{eq:scaling}
  x(c)\sim {1\over 2}-e^{-c}/2+c^2 e^{-2 c}/2.
\end{equation}

An interesting question is to ask the probability for a random
graph $G(n,c/(n-1))$ to have a perfect matching, a matching that
has $n/2$ edges. If the random graph
contains one or more isolated vertices, of cause there could be no
perfect matching. The average number of isolated vertices in a
random graph $G(n,c/(n-1))$ is $\langle n_0\rangle=n e^{-c}$. 
It must be less than unit for a perfect matching to be possible.
Therefore, $c\sim \ln(n)$. For $n=10^4$, $c\sim 9.21$.
Numerical experiment reveals that,
around $c\simeq 10$ there is a sharp change in the probability
of perfect matching for random graphs of size $n=10^4$. This is
in agreement with our above analysis.
For $c>10$  probability of perfect matching approaches unity while
for $c<10$ it approaches zero. Such a phenomenon was also 
observed in many other combinatorial optimization problems.
In the inset of fig.~\ref{fig:fig02} we demonstrate the 
experimental probability
for a random graph $G(n,c/(n-1))$ of $n=10^4$ 
to have a matching containing
at least $4900$ edges. 
We see a sharp transition occurs at $c\simeq 4.1$,
as would have been expected according to eq.~(\ref{eq:rsb}), which 
predicted that at $c=4.1$ the average max-matching size is $0.49 n$.
Figure~\ref{fig:fig02} also shows that the sharpness of this transition
is related to system size $n$, because the relative importance of
fluctuations in the max-matching size scales as $n^{-1/2}$.

\section{Conclusion}
To summarize, we have obtained an analytical expression for the
average max-matching size of a random graph as a function of the
average vertex degree $c$, and this formula was
verified by a numerical experiment. The analytical calculation was
performed on a designed spin model using the cavity method of statistical
physics. When $c>e$ the energy landscape of this model system 
has many valleys of the
same energy, separated by large energy barriers between them. The total
number of such low-energy valleys is also estimated. 
Different from the discontinuous glassy phase transition in
problems such as random 3-Sat, the transition in max-matching is
continuous, with a continuously growing structural entropy.

In a random graph, why there are great many maximum matching patterns
but only very few minimum vertex-covering 
patterns \cite{zhou2003b}? We think the reason is edge redundancy.
When the average vertex degree $c>e$, redundant edges give
additional freedoms to the max-matching patterns, but it
cause additional constraints to the min-covering patterns, as
all edges should be covered. 
As was shown here, when $c<e$
max-matching and min-covering are equivalent for random
graphs (in this sense, there is no redundant edges). 
Beyond $c=e$, there may still be deep connections between
the solution spaces of these two problems. An interesting
question is to investigate the possibility of constructing a
solution to the min-covering problem with the guide of the solutions
of the corresponding max-matching problem.

The excellent agreement between theory and experiment suggests that,
for short-range spin-glass models on finite connectivity random graphs,
the cavity field assumption that 
corresponds to the first-order replica-symmetry
breaking  might be accurate enough that no higher order RSBs is
needed for many purposes. For infinite connectivity SK model \cite{mezard1987}
and finite connectivity models on random graphs \cite{mezard2001,mezard2003}
we now enjoy satisfactory understandings. A challenge now
is the solution of spin glass 
models on finite connectivity regular lattices.

\acknowledgments
We are grateful to Professor YU Lu for support and
for revising  earlier versions of the manuscript. 
H.Z. acknowledges the hospitality of ITP and ICTS.

\section{Appendix: matching algorithm}
A max-matching pattern can be found with time proportional to polynomials of
graph size $n$ \cite{PapadimitriouC1998}. The algorithm mentioned below
is inspired by the concept of matching-alternating chain \cite{brualdi1999}. 
Suppose $M$ is a matching of a graph $(V,E)$. 
An incomplete matching-alternating chain (IMAC) of 
length zero is composed of a vertex that does not meet $M$; an IMAC
of length $p\geq 1$ is composed of a sequence of distinct vertices
$v_0, v_1,\ldots,v_p$ such that (1) $\{v_{i},v_{i+1}\}\in E$ for 
$i\in\{0,\ldots,p-1\}$;
(2) $v_0$ does not meet $M$ and $v_p$ meets $M$;
(3) the first, third,$\ldots$, edges do not belong to $M$;
and (4) the second, fourth,$\ldots$, edges belong to $M$.
Obviously, the length $p$ of an IMAC must be even. For
a complete matching-alternating chain (CMAC), $v_p$ also should
not meet $M$. The length of a CMAC must be odd. 

The algorithm works by inputting a matching $M$. It either returns
a new matching with size equaling to $|M|+1$ or stops if the input
matching is a max-matching.

(1). Find all the IMACs of length zero. Set all vertices as {\em unlabeled}.

(2). If there is no IMACs, stop. Else, 
select one IMAC (say chain $C_{v_a}$, which
is specified by its last 
element  $v_a$). Construct a 
set $A=\{v_i: V_i\;\;{\rm unlabeled}, \{v_a, v_i\}
\in (E-M)\}$.

(3). If $|A|\neq 0$,  select one of its elements (say $v_b$) and
delete $v_b$ from $A$. If $v_b$ is  already in chain $I_{v_a}$, return to 
step (3). Else,  mark $v_b$ as {\em labeled}. If $v_b$ does not meet $M$,
the chain formed by adding $v_b$ to the end of $C_{v_a}$ is a CMAC; jump to
(3a). Else, jump to (3b).

(3a).  Denote the set of edges in this CMAC as set $C$.
Denote $I=C\cap M$ as the set of edges which are shared by $C$ and $M$.
Then $M^\prime=(M-I)\cup(C-I)$ is a matching with of size $|M|+1$. 
Return $M^\prime$.

(3b). Suppose $v_c$ is the vertex such that 
$\{v_b, v_c\}\in M$. Create a new IMAC by adding 
$v_b$ and then $v_c$ to chain $C_{v_a}$.
Return to step (3).

(4). Delete chain $C_{v_a}$, and return to step (2).

If the algorithm returns a new matching, the algorithm is run again
with this new matching as input, till no new matching is returned.
It could proved that the last returned matching is a max-matching.

For each given value of $c$, we generate $1000-3000$ instances
for a random graph $G(n,c/(n-1))$ of order $n\sim 10^4$. 
The max-matching size for each of them
is obtained by the above-mentioned algorithm. 
The average max-matching sizes are shown in fig.~\ref{fig:fig02}
(filled circles).

\end{document}